\documentclass[amsmath,amssymb,twocolumn,aps,prd,10pt]{revtex4-2}
\usepackage[T1]{fontenc}
\usepackage{graphicx}
\usepackage{subfig}
\usepackage{epsfig}
\usepackage{dcolumn}
\usepackage{multirow}
\usepackage{rotating}
\usepackage{verbatim}
\usepackage{bm}
\usepackage{color}
\usepackage{soul}
\usepackage{float}
\usepackage{hyperref}

\def\lsim{\raise0.3ex\hbox{$<$\kern-0.75em\raise-1.1ex\hbox{$\sim$}}}
\def\gsim{\raise0.3ex\hbox{$>$\kern-0.75em\raise-1.1ex\hbox{$\sim$}}}

\def\pom{{I\!\!P}}

\def\beqa{\begin{eqnarray}}
\def\eeqa{\end{eqnarray}}

\begin{document}

\title{Exclusive dilepton production via timelike Compton scattering in heavy ion collisions }
\author{G. M. Peccini} 
\email{guilherme.peccini@ufrgs.br}
\author{L. S. Moriggi}
\email{lucas.moriggi@ufrgs.br}
\author{M. V. T. Machado}
\email{magnus@if.ufrgs.br}
\affiliation{High Energy Physics Phenomenology Group, GFPAE. Universidade Federal do Rio Grande do Sul (UFRGS)\\
Caixa Postal 15051, CEP 91501-970, Porto Alegre, RS, Brazil}

\begin{abstract}
In the present work, $k_T$-factorization formalism is applied to compute the exclusive dilepton production by timelike Compton scattering (TCS) in $eA$, $pA$ and $AA$ collisions. The nuclear effects are investigated considering heavy and light ions. The production cross section in terms of invariant mass and rapidity distribution of the lepton pair is shown. The analysis is done for electron-ion collisions at the Large  Hadron-Electron Collider (LHeC), its high-energy upgrade (HE-LHeC) and at the Future Circular Collider (FCC) in lepton-hadron mode. Additionally, ultraperipheral heavy ion collisions at future runs of the Large Hadron Collider (LHC) and at the FCC (hadron-hadron mode) are also considered.

\end{abstract}

\maketitle

\section{Introduction} 

Timelike Compton scattering (TCS) has been recently investigated in Ref.~\cite{Peccini:2020jkj} in the context of the $k_T$--factorization formalism. There, dilepton production was considered within a large range of dilepton invariant masses for the cases of electron-proton and proton-proton collisions. The calculation was based on Refs.~\cite{Schafer:2010ud,Kubasiak:2011xs}, where the process was studied for the first time. In this work, the aim is to extend that analysis considering nuclei as targets rather than protons. One interesting process is the electron-nucleus collision, which is planned to be investigated at the Electron Ion Collider (EIC) \cite{Accardi:2012qut} and at the Large Hadron-Electron Collider (LHeC) \cite{AbelleiraFernandez:2012cc}. Photonuclear reactions can be also studied in ultraperipheral heavy ion collisions \cite{Klein:2020fmr,Schafer:2020bnm} and it would be timely to analyze TCS in electromagnetic processes for large impact parameter proton-nucleus and nucleus-nucleus collisions.

Dilepton production can happen via different mechanisms. Two of the leading contributions come from Drell-Yan process and photon fusion, $\gamma \gamma \to \ell^+ \ell^-$. There is also a contribution from $\gamma \pom$ reaction in which the TCS provides the source of dileptons. Most of TCS studies so far have applied the formalism of Generalized Parton Distributions (GPDs) \cite{Diehl:2003ny,doi:10.1146/annurev.nucl.54.070103.181302,Belitsky:2005qn} (see also Refs.~\cite{Boer:2015fwa,Berger:2001xd,Moutarde:2013qs,Pire:2011st}). TCS amplitudes and related observables have been recently looked into using leading-twist approximation \cite{Grocholski:2019pqj} in the GPD approach. Further, investigations have been performed in order to lower the intrinsic model dependence. In Ref.~\cite{Lansberg:2015kha}, dilepton production through the TCS process was addressed in the context of ultraperipheral collisions (UPCs)  at a fixed-target experiment (AFTER@LHC), which was carried out utilizing the nucleon and ion beams.  

Timelike Compton scattering is the "opposite" process of deeply virtual Compton scattering (DVCS) in the sense that in TCS one has, at the final state, a virtual photon $\gamma p(A) \to \gamma^* p(A)$, whereas in DVCS there is a real photon, $\gamma^* p(A) \to \gamma p(A)$. The former has been carefully studied in Refs.~\cite{Machado:2008zv,Mariotto:2013qsa} for nuclear targets within the dipole formalism in the case of coherent scattering. The referred works were based on a previous paper concerning nuclear DVCS \cite{Machado:2008tp} and considered the space-like approximation. Predictions were presented for electron-ion collisions based on geometric scaling arguments. In Ref.~\cite{Motyka:2008ac} it was verified that in order to calculate the TCS cross section, one needs to deal with a strongly oscillatory integrand in the color dipole approach. To resolve this issue, it is necessary to make use of an analytic continuation of the integrand on the dipole size, $r$, and integrate it in the complex plane. Such procedure brought numerical difficulties for the calculations. These shortcomings do not appear in the momentum space and this is one of reasons to employ it in the present work. 

In the $k_T$-factorization framework, the gluon distribution depends on $x$ and $k_T^2$, where $k_T$ is the transverse momentum  of the corresponding parton. Formally, this distribution is called unintegrated Parton Distribution Function (uPDF). When the uPDF is integrated over $k_T$ with $Q^2$ being the upper limit, the usual Parton Distribution Function (PDF) is recovered. It turns out that in the small-$x$ regime gluons dominate and we simply designate the uPDF as UGD (unintegrated gluon distribution). At high energies, the $k_T$-factorization is a suitable formalism to compute the relevant distribution and cross sections. Within this regime, the longitudinal momentum fraction of partons, $x$, is small. This work is complementary to our previous study on electron-proton and proton-proton collisions. To adapt our treatment performed in \cite{Peccini:2020jkj} to nuclei targets, we replace the proton UGD by the nuclear one, applying the Glauber-Mueller formalism \cite{Glauber:1955qq,Mueller:1989st} to introduce the nuclear effects. The goal here is to examine TCS in nuclear targets for the first time in the $k_T$-factorization approach. In addition, it opens the possibility of carrying out a detailed study on the role played by the nonlinear QCD effects as the saturation scale is enhanced in nuclei in comparison with proton targets \cite{Peccini:2020tpj}.

The main goal is to investigate the nuclear effects in nuclear TCS, focusing on the atomic mass number ($A$) dependence. As in Ref.~\cite{Peccini:2020jkj}, we compute the cross section in terms of the dilepton invariant mass and rapidity at the center-of-mass energies of current and future machines for different nuclei. We are aware about the limitation on the use of factorization in nuclear collisions. In Refs.~\cite{Nikolaev:2003zf,Nikolaev:2004cu,Nikolaev:2005qs,Nikolaev:2006za} the validity of $k_T$-factorization for nuclear reactions was investigated and it was shown that linear $k_T$-factorization is broken in nuclear processes. In this context, it should be stressed out that this work is an exploratory study and further investigations should be carried out. 

This work is organized as follows. In Section II, we present the model from which we will build the nuclear unintegrated distribution function applying the Glauber-Mueller approach and discuss the technique employed to do so. In addition, we recall the main expressions for TCS calculation in proton targets and show how to adapt it to nuclear ones. We analyse the TCS cross section in UPCs in $pA$ and $AA$ collisions using the  equivalent photon approximation. Section III presents the numerical results and discussions for TCS observables in $eA$, $pA$ and $AA$ collisions. In Section IV, we summarize the results and conclusions.

\section{Dilepton production by nuclear TCS}

In the $k_T$-factorization the main ingredient is the UGD. Before presenting the TCS amplitude and cross section, in next subsection we briefly discuss and motivate the phenomenological model considered in the calculations: the Moriggi-Peccini-Machado (MPM) model \cite{Moriggi:2020zbv}.

\subsection{The MPM model}

The TCS process for $ep$ and $pp$ collisions has been previously calculated in detail in Ref.~\cite{Peccini:2020jkj}. The aim here is to extend that analysis to the case of nuclear targets. Within the $k_T$-factorization formalism, one can compute nuclear TCS cross section  by considering a nuclear UGD instead of the proton one. There are quite a few phenomenological models for the nuclear UGD \cite{Armesto:2002ny,Betemps:2009da,deOliveira:2013oma,Tribedy:2011aa,Dumitru:2018gjm}. On the other hand, it was also demonstrated \cite{Peccini:2020jkj} that distinct UGDs present practically the same results in the kinematical region we are taking into consideration. Thereby, we will focus on the MPM model, which was previously proposed by the authors in \cite{Moriggi:2020zbv}.

Briefly recalling this model, it is based on geometric scaling framework and defines an  expression for the gluon unintegrated function that depends on the variable $\tau$, being $\tau=k_T^2 / Q_s^2$, where $k_T^2$ is the transverse momentum squared of the gluons and $Q_s^2$ is the saturation scale. Alongside the quantity $\tau$, the MPM contains also three other parameters (see  \cite{Moriggi:2020zbv} for a detailed analysis). The distribution is given by
\begin{equation}
\phi_{\mathrm{MPM}} (x,k_T^2)= \frac{3 \sigma_0}{4 \pi ^2 \alpha_s}    \frac{(1+\delta n)}{Q_s^2}\frac{k_T^2}{ (1+ \frac{k_T^2}{Q_s^2})^{2+\delta n}},
\label{eq:mpm}
\end{equation}
in which $\delta n = a \tau ^b$ and $Q_s^2=(x_0/x)^{0.33}$. In the  expression above, $\sigma_0$, $x_0$, $a$ and $b$ were fitted against DIS (Deep Inelastic Scattering) data in the kinematic domain $x<0.01$. Beside describing DIS data at small-$x$, it also drives the spectra of produced hadrons in $pp/p\bar{p}$ processes. This model was built by means of the geometric scaling approach and a Tsallis-like behavior of the measured spectra. Additionally, the strong coupling does not depend on the scale $\mu^2$ and thus the constant value $\alpha_s=0.2$ is used.

The idea here is to adapt the MPM expression of the proton target, Eq.~(\ref{eq:mpm}), for the case of nuclear targets, and this can be conveniently performed by applying the technique utilized in \cite{Armesto:2002ny}, where the GBW UGD is extended to nuclei by using the Glauber--Mueller formalism. Following \cite{Armesto:2002ny}, the dipole scattering matrix in position space, $r$, is determined through the cross section for dipole-proton scattering,
\begin{equation}
\label{eq:SdA}
S_{dA}(x,r,b)=\mathrm{e}^{-\frac{1}{2} T_A(b)\sigma_{dp}(x,r)} \ .  
\end{equation}
The function $T_A(b)$ is the thickness function and depends on the impact parameter, $b$. Similarly to \cite{Armesto:2002ny}, we will assume a Woods--Saxon-like parametrization for the nuclear density \cite{DEVRIES1987495} (except for Li, for which the nuclear density will be taken as Gaussian distribution) whose normalization is $\int d^2b T_A(b) = A $. Thus, the nuclear UGD is written in the following way:
\begin{equation}
\label{eq:fiA}
\phi_A(x,k_T^2,b)=\frac{3}{4\pi^2\alpha_s}k_T^2\nabla^2_{k_T} \mathcal{H}_0 \left\{ \frac{1-S_{dA}(x,r,b)}{r^2}\right\},
\end{equation}
where $\mathcal{H}_0 \left\{ f(r) \right\}=\int dr\, r J_0(k_Tr)f(r)$ is the order zero Hankel transform.  

Regarding the proton target, a homogeneous object with radius $R_p$ is considered, which factorizes  
$S_{dp}(x,r,b)$ into $S_{dp}(x,r,b)=S_{dp}(x,r)\Theta (R_p-b)$. For large dipoles, $S_{dp}(x,r) \rightarrow 0$, and the cross section reaches a bound given by $\sigma_0=2\pi R_p^2$. In the saturation approach, the gluon distribution presents a maximum at $k_T \simeq Q_s(x)$. This formalism is characterized by geometric scaling, which implies that observables become dependent on the ratio $Q^2/Q^2_s(x)$ instead of $Q^2$ and $x$ separately.

The dipole cross section in coordinate space $r$ may be evaluated as \cite{Moriggi:2020zbv},
\begin{eqnarray}
\label{eq:DISc}
\sigma_{dp}(\tau_r)=\sigma_0\left (  1-\frac{2(\frac{\tau_r}{2})^{\xi}K_{\xi}(\tau_r)}{\Gamma(\xi)} \right ),
\end{eqnarray}
where $\xi =1+\delta n$ and $\tau_r = rQ_s(x)$ is the scaling variable in the position space. Accordingly, the nuclear gluon distribution is obtained from Eqs. (\ref{eq:SdA}) and (\ref{eq:fiA}).

At this point, some considerations are in order. The shadowing of structure functions observed in nuclear DIS in the region of small-$x$ is viewed in the saturation formalism/CGC as the multiple scattering of the photon fluctuations in the nuclear target, giving rise to the modification of nuclear UGD compared to that of free nucleons. It is well known that this effect is enhanced as the atomic mass number, $A$, increases \cite{Arneodo:1996rv}. Here, we investigate the $A$-dependence of the cross section for diffractive production of dileptons at current and future colliders (HL-LHC/LHeC, HE-LHC/HE-LHeC and FCC-eA/pA(AA)). At high energies the small-$x$ regime is  reached, $x\sim M^2/W^2 \lesssim 10^{-6}$, where it is expected a significant suppression of this observable relative to the $ep$ case. The nuclear structure functions at small-$x$ were constrained experimentally by E665 and NMC collaborations for the nuclei Li, C, Ca, Sn and Pb \cite{Adams:1995is,Arneodo:1995cs,Arneodo:1996ru,Arneodo:1996rv}. In our analysis, we utilize these nuclei as representative targets and carry out predictions for nuclear TCS. In next subsection, the expression for TCS amplitude is reviewed and the photonuclear case is also discussed.

\subsection{TCS in electron-nucleus collisions}

In what follows, we will recall the main expressions in \cite{Peccini:2020jkj} about TCS in electron-proton collisions. Therein, it was shown that the imaginary part of the TCS amplitude is written as
\begin{equation*}
\label{eq:ImAf}
\text{Im}\mathcal{A}_f^\text{TCS}= \frac{4  \alpha_{em}e_f^2}{\pi} \Big [ \Theta (M_{\ell^+\ell^-}^2 - 4m_f^2)
\end{equation*}
\begin{equation*}
\times \big (\text{PV} \int _{4m_f^2}^{\infty}\Omega (W^2, M_{q\bar{q}}^2,M_{\ell^+\ell^-}^2)  \ dM_{q\bar{q}}^2
\end{equation*}
\begin{equation*}
+\pi \text{Re}\mathcal{M}_f(W^2,M_{\ell^+\ell^-}^2)\big )+ \Theta (4m_f^2 - M_{\ell^+\ell^-}^2)
\end{equation*}
\begin{equation}
\times \int _{4m_f^2}^{\infty} \Omega (W^2,M_{q\bar{q}}^2,M_{\ell^+\ell^-}^2)  \ dM_{q\bar{q}}^2\Big ] \ .
\label{eq:theta}     
\end{equation}
Analogously, the real part is given by
\begin{equation*}
\text{Re}\mathcal{A}_f^\text{TCS}= \frac{4  \alpha_{em}e_f^2}{\pi} \Big [ \Theta (M_{\ell^+\ell^-}^2 - 4m_f^2)
\end{equation*}
\begin{equation*}
\times \big (\text{PV} \int _{4m_f^2}^{\infty}\eta (W^2,M_{q\bar{q}}^2,M_{\ell^+\ell^-}^2)  \ dM_{q\bar{q}}^2
\end{equation*}
\begin{equation*}
-\pi \text{Im}\,\mathcal{M}_f(W^2,M_{\ell^+\ell^-}^2)\big )
+ \Theta (4m_f^2 - M_{\ell^+\ell^-}^2) 
\end{equation*}
\begin{equation}
\times \int _{4m_f^2}^{\infty} \eta (W^2,M_{q\bar{q}}^2,M_{\ell^+\ell^-}^2)  \ dM_{q\bar{q}}^2\Big ]       \ .
\end{equation}

The definitions of $\Omega (W^2,M_{q\bar{q}}^2,M_{\ell^+\ell^-}^2)$ and $\eta (W^2,M_{q\bar{q}}^2,M_{\ell^+\ell^-}^2)$ are the following:
\begin{equation}
\Omega (W^2,M_{q\bar{q}}^2,M_{\ell^+\ell^-}^2) =  \frac{\text{Im}\mathcal{M}_f (W^2,M_{q\bar{q}}^2)}{M_{q\bar{q}}^2 - M_{\ell^+\ell^-}^2} \ ,
\end{equation}
\begin{equation}
\eta (W^2,M_{q\bar{q}}^2,M_{\ell^+\ell^-}^2) =  \frac{\text{Re}\mathcal{M}_f (W^2,M_{q\bar{q}}^2)}{M_{q\bar{q}}^2 - M_{\ell^+\ell^-}^2} \ . 
\end{equation}
In the previous expressions, $e_f$ is the quark charge of flavor $f$, while $m_f$ is its mass. The quantities $W$, $M_{q\bar{q}}^2$ and $M_{\ell^+\ell^-}^2$ are the photon-nucleus center-of-mass energy, dipole invariant mass squared and dilepton invariant mass squared, respectively. For further details on the expressions, see Ref.~\cite{Peccini:2020jkj}.

\begin{figure*}[t]
\centering
    \includegraphics[width=0.8\textwidth]{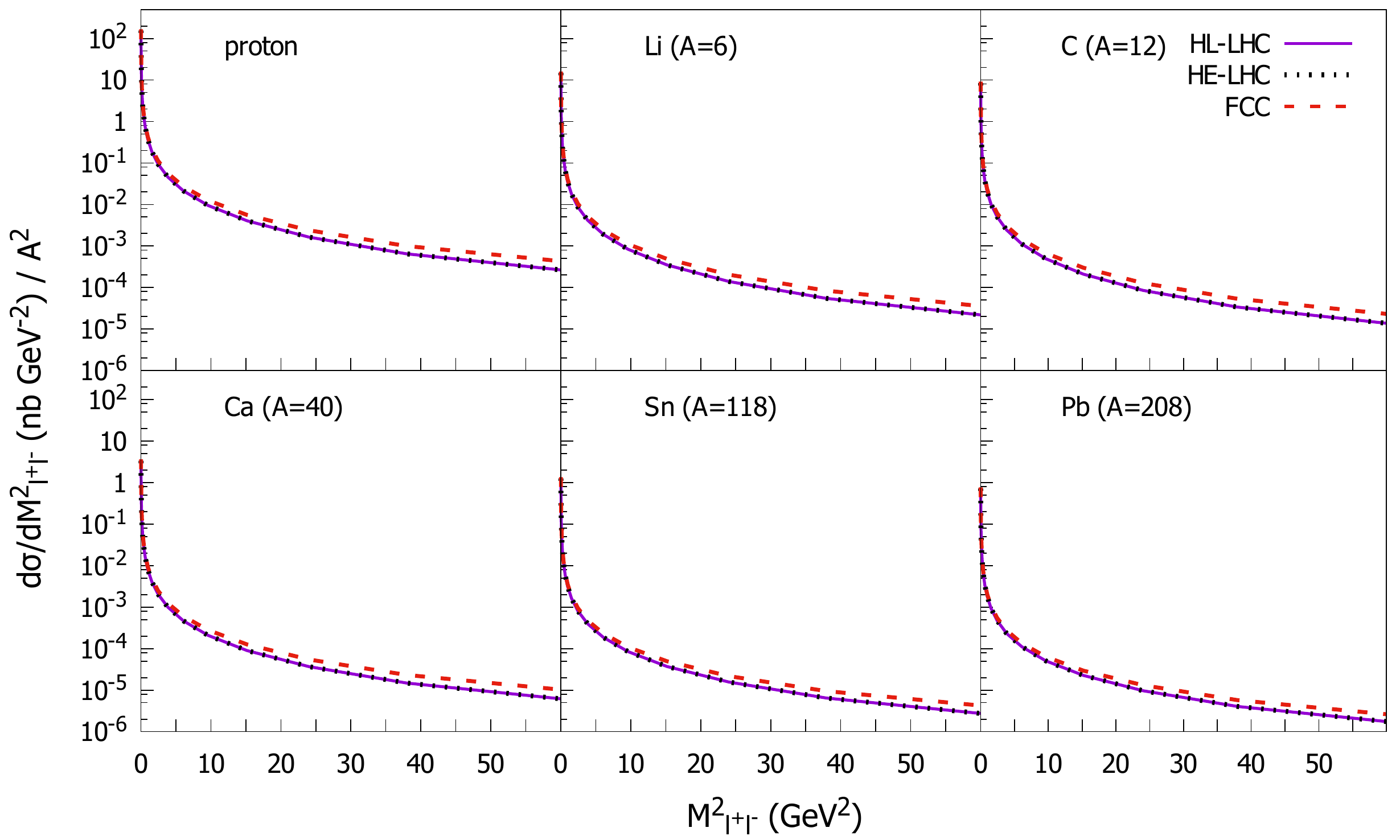}
    \caption{Differential cross section for dilepton production  as a function of dilepton invariant mass in $eA$ collisions calculated through Eq.~(\ref{eq:difXsec}) for different nuclei at the energies described in Table \ref{tab:2}.}
    \label{fig:dsigdM2eA}
\end{figure*}

Adopting the dipole picture, where the virtual photon fluctuates into a quark-antiquark pair, the spectral distribution in Eq. (\ref{eq:theta}) that is related to the diffractive amplitude for the $\gamma A \to q\bar{q}A$ transition is given by
\begin{eqnarray}
\text{Im}\,\mathcal{M}_f(W^2,M_{q\bar{q}}^2)&=&\frac{1}{\pi M_{q\bar{q}}^2}\int _0 ^{\kappa_{\mathrm{max}}^2}\frac{d^2\kappa}{\sqrt{1-4\left(\frac{\alpha}{M_{q\bar{q}}^2}\right)}} \nonumber \\
&\times & \text{Im}\,M_f(W^2,\kappa^2,z)  \ , 
\label{eq:spectral}
\end{eqnarray}
where $\kappa_{\mathrm{max}}^2 = (0.25M_{q\bar{q}}^2-m_f^2)$ and 
\begin{equation}
\text{Im}M_f(W^2,\kappa^2,z) = \int _0 ^ \infty \frac{d^2 k_{\perp}}{k_{\perp}^2} \phi_A(x,k_{\perp}^2) \alpha_s(\mu ^2)
\end{equation}
\begin{equation}
\times \big [C_{0f}(z,\kappa^2)D_{0f}(\kappa^2,k_{\perp}^2)+C_{1f}(z,\kappa^2)D_{1f}(\kappa^2,k_{\perp}^2)\big ] \ .
\label{eq:main}
\end{equation}
The functions $C_{0f}$, $D_{0f}$, $C_{1f}$ and $D_{1f}$ are specified in \cite{Peccini:2020jkj}. Recalling the determination of $\text{Re}M_f$, it is computed via the dispersion relation,
$\rho = \text{Re}\,M_f/\text{Im}M_f$. The $\rho$ parameter is defined as $\rho = \tan \left( \frac{\pi}{2} \lambda_{\mathrm{eff}}\right)$, in which $\lambda_{\mathrm{eff}}=\partial \ln(\text{Im} M_f)/\partial \ln(W^2)$.

In order to embed a $t$ dependence in the scattering amplitude, one needs to take into consideration the nuclear form factor. For simplicity and following Ref.~\cite{Klein:1999qj}, an analytical expression based on homogeneous hard sphere and the Yukawa approximation will be considered:
\begin{eqnarray}
\label{eq:hardsphere}
F(q)  =  \frac{4 \pi \rho_0}{A q^3} \big [ \sin(q R_A) - qR_A \cos (qR_A)   \big ] \nonumber \\
\times \big [ \frac{1}{1+r_0^2q^2}\big ]
\label{eq:formfactor}
\end{eqnarray}
where $q=\sqrt{|t|}$, $\rho_0=A/(4/3 \pi R_A^3)$ and $R_A=(1.12 A^{1/3}-0.86 A^{-1/3})$. The parameter $r_0$ is the range of a Yukawa potential and its value is $r_0=0.7$\,fm. Therefore, the amplitude depending on $t$ is expressed as follows:
\begin{equation}
A_f^\text{TCS} (W,t)= F(q) A_f^\text{TCS}(W,t=0)   
\end{equation}
The differential cross section  for the $\gamma A \to \gamma^{*}A $ collision is then
\begin{eqnarray}
\frac{d \sigma}{dt}(\gamma A \to \gamma ^* A )  =  \frac{[\text{Im} (A^\text{TCS})]^2 \left(1+\rho^2 \right)}{16 \pi } |F(q)|^2,
\end{eqnarray}
where $\text{Im} A^\text{TCS} = \sum \text{Im} A^\text{TCS}_f$ , with the summation over quark flavour. The integrated cross section is given by
\begin{eqnarray}
\sigma (\gamma A \to \gamma ^* A ) = \left. \frac{d\sigma }{dt}\right|_{t=0} \int_{-\infty}^{t_{min}}|F(q)|^2 dt  ,
\end{eqnarray}
Having the TCS cross section, one may express the differential cross section in terms of the dilepton invariant mass distribution, 
\begin{equation}
\frac{d \sigma (\gamma A \to \ell^+\ell^- A)}{dM_{\ell^+\ell^-}^2} = \frac{\alpha_{em}}{3 \pi M_{\ell^+\ell^-}^2} \sigma^\text{TCS}(\gamma A \to \gamma ^* A )  .
\label{eq:difXsec}
\end{equation}

\subsection{TCS in ultrapheripheral heavy ion collisions}

\begin{figure*}[t]
\centering
    \includegraphics[width=0.8\textwidth]{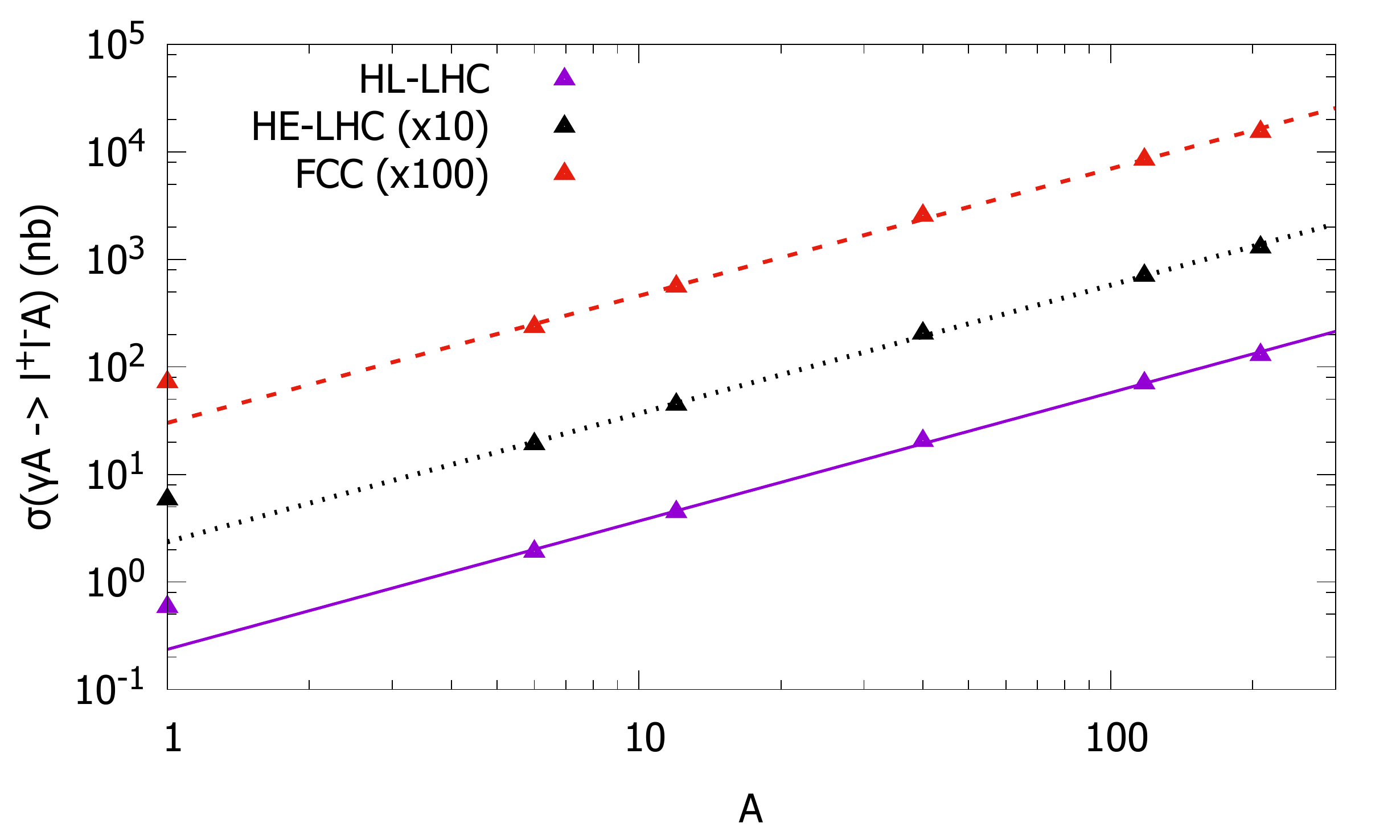}
    \caption{Integrated cross section in $eA$ collisions as a function of the atomic mass number. Each line is a power-like fit, $A^{\alpha}$, with $\alpha =1.19$.}
    \label{fig:eAtotal}
\end{figure*}

Photonuclear reactions can be also investigated in ultra relativistic heavy ion collisions in the case of large impact parameter. The protons or heavy nuclei are then sources of quasi-real photons. For $pA$ collisions, analogously to the $pp$ case (see Ref.~\cite{Peccini:2020jkj}), dilepton production through TCS is dominated  by $\gamma \pom$ and $\pom \gamma$ reactions at high energies (an analysis on ultraperipheral TCS is performed in Ref. \cite{Pire:2008ea}). Within the equivalent photon approximation (EPA), the differential cross section for the nuclear coherent scattering, $p+A \to p+ \ell^+ \ell ^-+A$, in terms of dilepton invariant mass and rapidity, is given by \cite{Lansberg:2015kha} 
\begin{eqnarray}
\frac{d\sigma^{pA} }{d M^2_{\ell^+
\ell^-} dy_{\mathrm{pair}}} & = & k_{+}\bigg [  \frac{dN (k_+)}{dk_+} \bigg ]_{p} \ \bigg [ \frac{d \sigma ^{\gamma A \to \ell^+\ell^- A}}{dM^2_{\ell^+\ell^-}}  (W_+) \bigg ]  \nonumber \\
& + & k_-    \bigg [ \frac{d N(k_-)}{dk_-} \bigg ]_{A}  \ \bigg [ \frac{d\sigma^{\gamma p \to \ell^+\ell^- p}}{dM^2_{\ell^+\ell^-}} (W_-) \bigg ],
\end{eqnarray}
In the  expression above,  $k$ stands for the photon energy, $dN (k)/dk$ is the photon flux and $y_{pair}$ is the rapidity of the lepton pair. 

The photon flux of the proton will be given by \cite{Drees:1988pp},
\begin{equation*}
\bigg [ \frac{d N(k)}{dk}\bigg ]_{p}=\frac{\alpha_{em}}{2 \pi k} \bigg [1+\bigg (1-\frac{2k}{\sqrt{s}} \bigg )^2\bigg ]
\end{equation*}
\begin{equation}
\times \bigg ( ln \ \chi  -\frac{11}{6}+\frac{3}{\chi}-\frac{3}{2 \chi^2}+\frac{1}{3\chi^3} \bigg )  \ .
\end{equation}
in which $\chi=1+(Q_0^2/Q_{min}^2)$ with $Q_0^2 =0.71 \ GeV^2$ and $Q_{min}^2=k^2 / \gamma_L^2$, where $\gamma _L=\sqrt{s}/2m_p$. For nuclei, the photon flux is written as follows \cite{Bertulani:2005ru}: 
\begin{eqnarray}
\bigg [ \frac{dN(k)}{dk} \bigg ]_A & = & \frac{2Z^2 \alpha_{em}}{\pi k}
\bigg [ \Delta K_0 (\Delta) K_1(\Delta)  \nonumber \\ 
& - & \frac{\Delta ^2}{2} \big (K_1^2 (\Delta) - K_0^2(\Delta) \big ) \bigg ] 
\label{eq:nuclearflux}
\end{eqnarray}
where $\Delta = 2 k R_A / \gamma_L$ in $AA$ collisions and $\Delta \approx  k (R_p+R_A) / \gamma_L$ for $pA$ collisions.
 
The photon energy $k$ and the center-of-mass energy $W$ of the photon-nucleus (nucleon) can be written in terms of the dilepton rapidity and its invariant mass,
\begin{eqnarray}
k_{\pm}=\frac{M_{\ell^+\ell^-}}{2}  e^{\pm y_{\mathrm{pair}}}, \quad W^2_{\pm}= 2 k_{\pm} \sqrt{s} .
\end{eqnarray}
The expressions above relate the photon-proton center-of-mass energy to the proton-nucleus and nucleus-nucleus ones.
In case of $AA$ collisions, the differential cross section takes the following form:
\begin{eqnarray}
\frac{d\sigma^{AA} }{d M^2_{\ell^+
\ell^-} dy_{\mathrm{pair}}} & = & k_{+}\bigg [  \frac{dN (k_+)}{dk_+} \bigg ]_{A} \ \bigg [ \frac{d \sigma ^{\gamma A \to \ell^+\ell^- A}}{dM^2_{\ell^+\ell^-}} (W_+) \bigg ] \nonumber \\
& + & k_-    \bigg [ \frac{d N(k_-)}{dk_-} \bigg ]_{A}  \ \bigg [ \frac{d\sigma^{\gamma A \to \ell^+\ell^- A}}{dM^2_{\ell^+\ell^-}}(W_-) \bigg ],
\end{eqnarray}

In the next section we perform the numerical calculations for exclusive dilepton production in $eA$ and $p(A)A$ collisions.

\begin{figure*}[t]
\centering
    \includegraphics[width=0.8\textwidth]{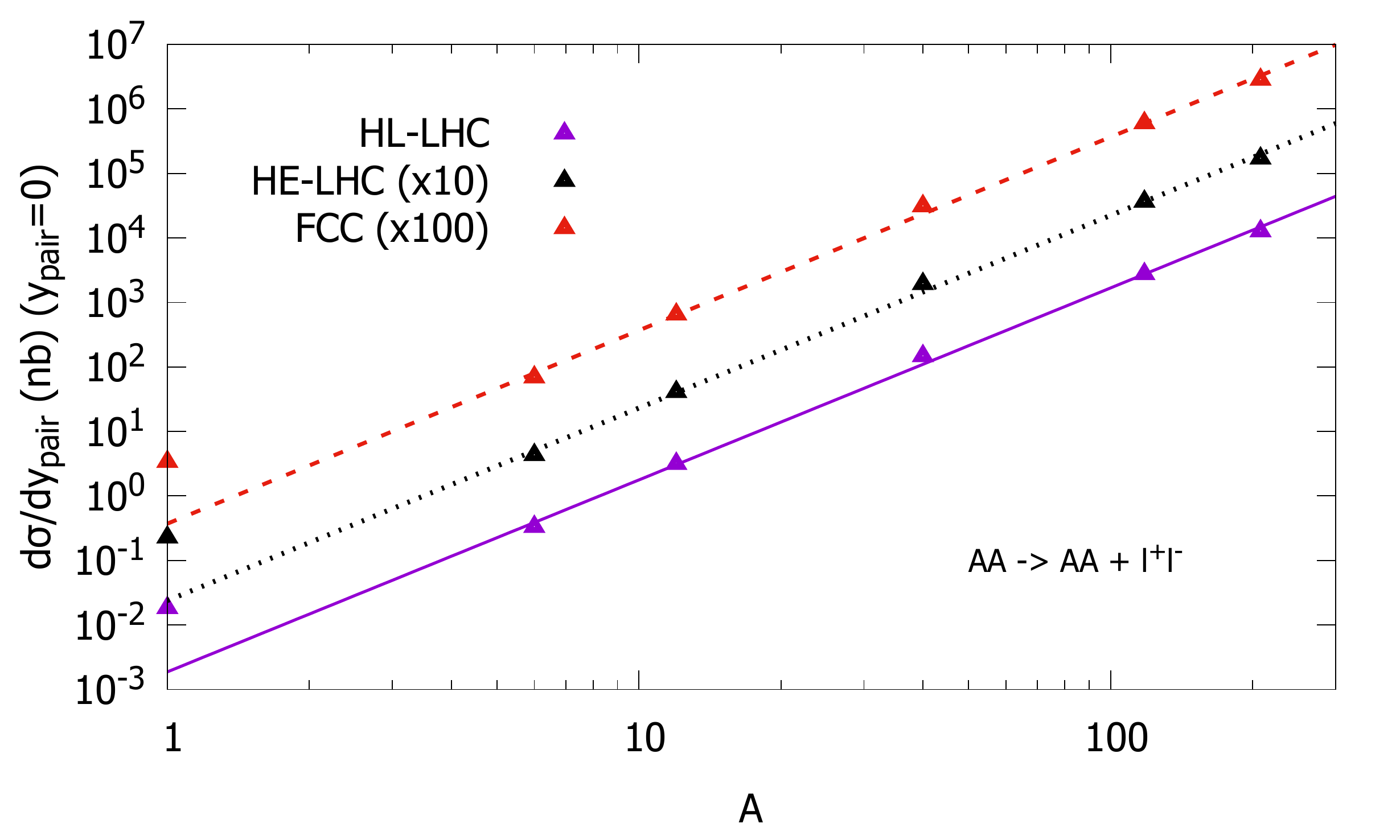}
    \caption{Cross section for $AA$ collisions, $d\sigma /dy_{pair} (AA)$, at $y_{pair}=0$ as a function of the atomic mass number. Each line is a power-like fit, $A^{\beta}$, with $\beta=2.9$.}
    \label{fig:DSDY0}
\end{figure*}

\begin{figure*}[t]
\centering
    \includegraphics[width=1.0\textwidth]{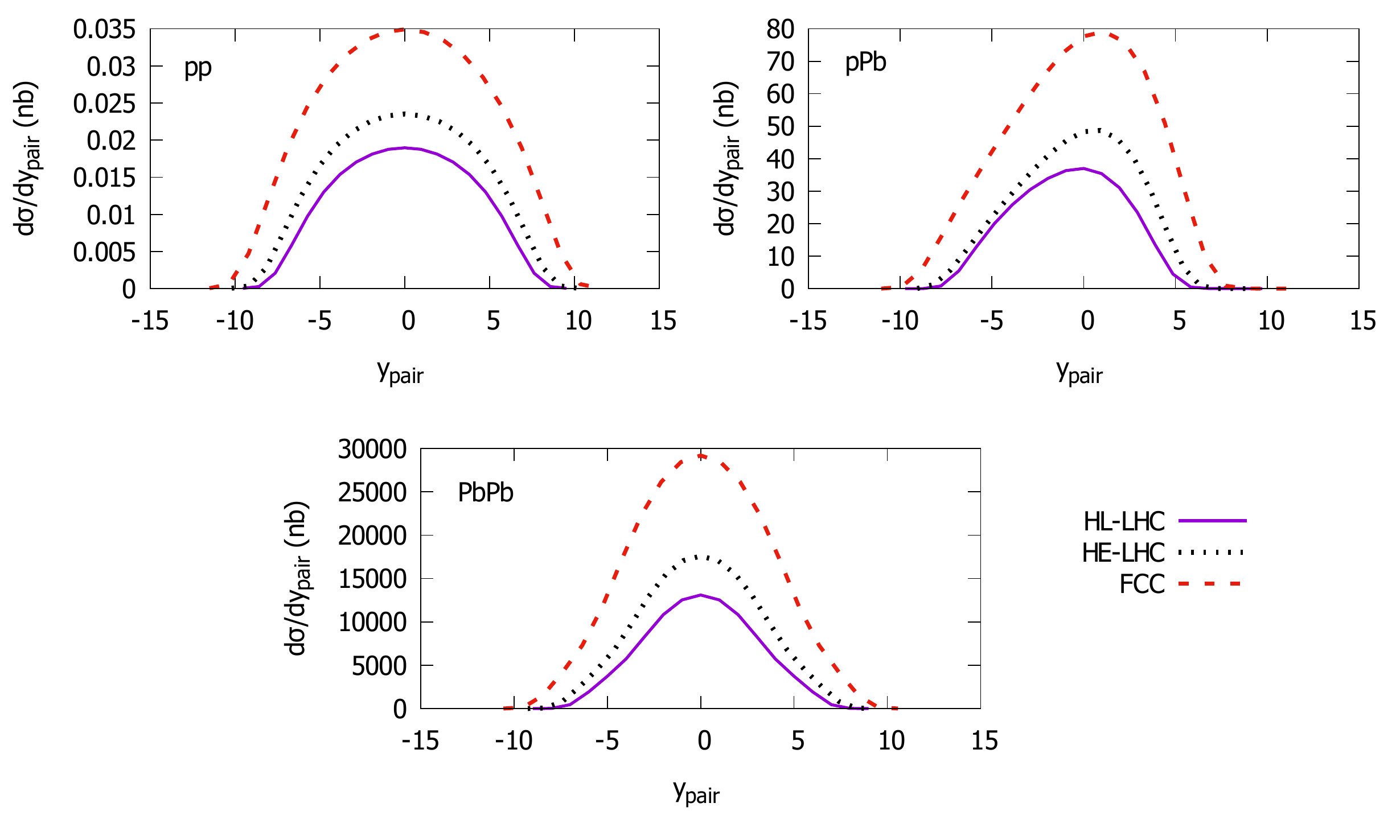}
    \caption{Predictions for rapidity distributions in $pp$, $pPb$ and $PbPb$ collisions at the energies of the considered colliders.}
    \label{fig:DSDY}
\end{figure*}

\section{Numerical results and discussion}

The TCS cross section will be computed in terms of dilepton invariant masses and rapidity distributions for $eA$, $pA$ and $AA$ collisions in the kinematic domain of current and future colliders. Their design configurations are summarized in Table \ref{tab:1}.
Different nuclei are considered (light and heavy ones) in order to cover the wide range on the atomic mass number, $A$. In that sense, the observables will be evaluated for Li (Z=3, A=6), C (Z=6, A=12), Ca (Z=20, A=40), Sn (Z=50, A=118) and Pb (Z=82, A=208). The center-of-mass energies for each nuclear collision are estimated  considering the energy of the proton beam for each machine. The energies of the proton and electron beams are based on the prospects in Ref.~\cite{Bordry:2018gri}.

\begin{table}[t]
\begin{center}
\begin{tabular}{|l|c|c|}
\hline

\textbf{Collider}  & $E_p$ & $E_e$ \\
\hline
LHeC/HL-LHeC (TeV)    & 7 & 0.06   \\
HE-LHeC (TeV)   & 13.5	& 0.06   \\
FCC-eh (TeV)  & 50 & 0.06  \\
\hline 
HL-LHC (TeV) &  7 & -  \\
HE-LHC (TeV) & 13.5 & - \\
FCC-hh (TeV) & 50 & - \\
\hline

\end{tabular}
\end{center}
\caption{Configurations of the projected beam energies for future  high energy machines \cite{Bordry:2018gri}.}
\label{tab:1}
\end{table}

First, we present the results for the electron-ion mode. Fig.~\ref{fig:dsigdM2eA} shows the differential cross section (divided by $A^2$) for dilepton production using Eq.~(\ref{eq:difXsec}) at the energies and atomic mass numbers presented in Table \ref{tab:2}. Namely, we present prediction for the future machines LHeC, using the high luminosity (HL) and high energy (HE) baseline parameters at the LHC. Also, the forthcoming Electron-Ion Collider (EIC) and the Future Circular Collider (FCC) in $eh$ mode were taken into account.  The typical monotonic decrease of cross section on the dilepton invariant mass appears and the low mass region is the dominant contribution. It is clearly seen that there is a small enhancement of the cross section at FCC energy relative to HL/HE-LHC.

\begin{table}[t]
\begin{center}
\begin{tabular}{|l|c|c|c|c|}
\hline
  & LHeC/HL-LHeC (TeV) & HE-LHeC (TeV) & FCC-eA (TeV) \\
\hline
Pb    & 0.81 & 1.13  & 2.18  \\
Sn   & 0.84	& 1.17 & 2.25	  \\
Ca     & 0.92 & 1.27  & 2.45 \\
C   & 0.92 & 1.27    &  2.45  \\
Li & 0.92 & 1.27    &  2.45 \\
\hline
\end{tabular}
\end{center}
\caption{Estimated center-of-mass energies (per nucleon) of future electron-ion machines given a nucleus beam.}
\label{tab:2}
\end{table}


One realizes that there is a significant decrease of the cross section as $A$ increases, which indicates the presence suppression due to nuclear effects. In the absence of such effects, one expects that the cross section scales with $A^2$ since $\sigma (\gamma A \rightarrow \gamma^* A) $ depends on the gluon distribution squared and the form factor of Eq.~(\ref{eq:formfactor}) gives a nuclear slope which depends, roughly speaking, on $R_A^2\sim A^{2/3}$. The nuclear shadowing content in the UGD should produce a decreasing of the integrated cross section in terms of $A$. One can parametrize the growth of the integrated cross section as a power law $A^{\alpha }$ (with $\alpha = 2-2/3-\delta$), where the parameter $\delta$ results from the nuclear shadowing. In Fig.~\ref{fig:eAtotal}, we plotted the TCS integrated  cross section for dilepton production as a function of the atomic mass number, $A$. Larger cross section occurs for the FCC - eA energy, as already expected. The lines represent a power law fit  $A^{\alpha}$ with $\alpha = 1.19$ and it seems energy independent. In that sense, the value of $\alpha$ is close to the expected one for weak absorption where it  grows as $A^{4/3}$ ($\alpha \approx 1.33$). Such a behavior is quite similar to the photonuclear production of heavy vector mesons as $J/\psi$ in the context of vector meson dominance (VMD) and small absorption in the Glauber model calculation. On the other hand, for strong absorption the expected behavior would be the black disk scaling, $\sigma_{\gamma A} \sim A^{2/3}$. For light nuclei the cross section has magnitude of units of nb and for heavy ions it reaches $\sim 100$\,nb.

\begin{table}[t]
\begin{center}
\begin{tabular}{|l|c|c|c|c|}
\hline
   & HL-LHC [TeV] & HE-LHC [TeV] & FCC-pA (AA) [TeV] \\
\hline
Pb      & 8.79 (5.52)  & 16.95 (10.64) & 62.79 (39.42)  \\
Sn      & 9.11 (5.93) & 17.58 (11.44) & 65.09 (42.37)	  \\
Ca       & 9.90 (7.00)  & 19.09 (13.50) & 70.71 (50.00) \\
C     & 9.90 (7.00)    &  19.09 (13.50) & 70.71 (50.00)  \\
Li    & 9.90 (7.00)    &  19.09 (13.50) & 70.71 (50.00) \\
\hline
\end{tabular}
\end{center}
\caption{Estimated center-of-mass energies (per nucleon) of pA (AA) collisions at the HL-LHC, HE-LHC and FCC-pA (AA) given a nucleus beam.}
\label{tab:3}
\end{table}


With regard to the predicted TCS integrated cross section for the upcoming Electron-Ion Collider (EIC), we only considered Au (gold) nucleus at collision energy of $\sqrt{s}=92$ GeV. The obtained value was $52.19$ nb and the event rate per year, given the EIC luminosity ($L=10^{33-34}$ cm$^{-2} $s$^{-1}$) \cite{Accardi:2012qut}, is predicted to be approximately $1.646 \times 10^{10}$. 

Lastly, we also accounted for the current configurations of the LHC (Run 2). The TCS integrated cross sections for pp ($\sqrt{s}=13$ TeV), pPb ($\sqrt{s}=8.16$ TeV) and PbPb ($\sqrt{s}=5.02$ TeV) are 106 pb, 151 nb and 48 $\mu$b, respectively. The corresponding event rates per year, assuming the LHC Run 2 luminosity ($L=10^{34}$ cm$^{-2}$ s$^{-1}$) \cite{Boyd:2020qox} are the following: $3.343 \times 10^{7}$, $4.762 \times 10^{10}$ and $1.514 \times 10^{13}$. 

The results for $eA$ collisions can be directly compared to the prediction using the color dipole picture. In Ref.~\cite{Machado:2008zv} the nuclear TCS cross section has been evaluated. In particular, it was considered a lead nucleus and two models for the dipole-nucleus cross section:  the Marquet-Peschanski-Soyez (MPS) \cite{Marquet:2007qa} and b-SAT \cite{Kowalski:2006hc} models. It was obtained $\sigma^\text{TCS} \simeq 15$ nb ($22$ nb) at the LHeC energy (W $\approx 800$ GeV), $\sigma^\text{TCS} \simeq 19$ \ nb ($23$ nb) at the HE-LHeC energy (W $\approx 1200$ GeV) and $\sigma^\text{TCS} \simeq 22$ nb  ($25$ nb) at the FCC-eA energy (W $\approx 2200$\,GeV), using the MPS (b-Sat) model, respectively. These values are consistent with ours for $A=208$ (see Fig.~\ref{fig:eAtotal}). There, the integrated  TCS cross section considers $M^2_{\ell^+
\ell^-}\geq 2.25$\,GeV$^2$, whereas in our work the lower bound of the integral was taken as $M^2_{\ell^+\ell^-}=1$\,GeV$^2$, thus the results in \cite{Machado:2008zv} should be smaller than ours. In addition, in Ref.\cite{Machado:2008zv} a spacelike approximation has been considered which provides cross sections smaller than the correct timelike kinematics (see discussion in Refs.~\cite{Motyka:2008ac,Schafer:2010ud}).

We now turn to the ultraperipheral heavy ion collisions. The collision energies taken into consideration are outlined in Table \ref{tab:3}. In Fig.~\ref{fig:DSDY0}, the dilepton rapidity distribution is shown for $y=0$. Due to $Z^2\sim A^2$, the leading small $\Delta$ (equivalently, small $k$) contribution in Eq.~(\ref{eq:nuclearflux}) and the fact that we obtained $\sigma(\gamma A \to \ell^+ \ell^- A $) scaling with $A$ as $\sigma(\gamma A \to \ell^+ \ell^- A ) \sim A^{1.19}$, roughly one should have $d\sigma_{AA}/dy\sim A^{3.19}$. Despite the approximations, we got $d\sigma_{AA}/dy\sim A^{\beta}$  (with $\beta=2.9$), which is close to the theoretical expected  value. Concerning the typical order of magnitude, for lead one reaches dozens of $\mu$b at midrapidities. This is translated into a event rate for TCS channel of $\sim 10^5$ Hz at the HL-LHC. 

The separation of different channels for the dilepton production in $pp$ case  involves the investigation of correlations for the outgoing particles, as the transverse momenta of final state protons or outgoing muons. Correlations in the rapidity space for outgoing leptons  can be also considered. It was shown in \cite{Kubasiak:2011xs} that correlations for
the (single, double) diffractive mechanism are more intense than that for the two photon fusion, and the exclusive dilepton (TCS) production has same order of magnitude than the central diffractive one. The same selection can be considered in $pA$ or $AA$ collisions. The low $p_T$ region is of great interest and in Ref.~\cite{Klein:2020jom} a careful study has been done in this kinematic domain. It includes the initial contributions due to the incoming photons, the  soft photon radiations expressed in a Sudakov resummation, the  multiple interactions between the lepton pairs and the electric charges inside the QGP and  the effects of an external magnetic fields.

In Fig.~\ref{fig:DSDY}, the rapidity distribution is shown for $pp$, $pPb$ and $PbPb$ collisions. The cross sections considerably increase for higher energies. As expected, the $pPb$ case  has an asymmetric rapidity distribution, contrary to  $pp$ and $PbPb$ collisions. In Table \ref{tab:4}, numerical estimates for the integrated TCS total cross section are presented  for these  three collision modes. The values slightly differ from one machine configuration to another. The results for $pp$ are quite similar to that one presented in our previous work \cite{Peccini:2020jkj}, where the theoretical uncertainty coming from the choice of the UGD has been discussed. For PbPb collisions, the TCS channel can be compared with the exclusive production of dilepton from two photon fusion. Recently, in Ref.~\cite{Azevedo:2019fyz} this channel has been analysed for the energy of $\sqrt{s_{NN}}=5.02$ TeV at the LHC and it was found $d\sigma/dy (y=0)\simeq 250-300$ mb without any cut (the band corresponds to different treatments for the photon luminosity). Afterwards, in Ref.~\cite{Goncalves:2020btj} the background associated with the diffractive production was investigated and it has been shown that such a channel is strongly suppressed. A new analysis for the exclusive dilepton production in $\gamma \gamma$ reaction in UPCs has been presented in \cite{Klusek-Gawenda:2020eja}, where the differential cross section is computed using the complete photon´s polarization
density matrix resulting from the Wigner distribution framework. The authors claim that the approach provides much better agreement with experimental data than other approaches available in the literature. In the proton-lead collisions, the prediction can be compared to the recent analysis in Ref.~\cite{Dyndal:2019ylt} where a new experimental method to probe the photon PDF inside the proton at the LHC has been proposed. Interestingly, an unintegrated photon distribution (photon UGD) was considered and it was shown that due to the smearing of dilepton $p_T$ introduced by the $k_T$-factorization formalism, the cross section is about 1/3 higher than the expected from usual collinear factorization. It would be timely to impose the same cuts in order to understandd the background coming from the TCS process.

\begin{table}[t]
\begin{center}
\begin{tabular}{|l|c|c|c|c|}
\hline
  & HL-LHC  & HE-LHC  & FCC-hh  \\
\hline
pp     & 0.110 (nb) &  0.137 (nb)  & 0.206 (nb) \\
pPb   & 0.155 ($\mu$b)  & 0.228 ($\mu$b)  & 0.360 ($\mu$b) 	  \\
PbPb      & 0.0500 (mb) & 0.0787 (mb)  & 0.1335 (mb)  \\

\hline
\end{tabular}
\end{center}
\caption{Integrated cross sections for $pp$, $pPb$ and $PbPb$ collisions at different configurations at the LHC ($pp$ and heavy ion modes). }
\label{tab:4}
\end{table}

Finally, concerning the predicted breaking of $k_T$-factorization  mentioned already, some discussion is needed. The factorization theorems in perturbative QCD (pQCD) have the fundamental point that hard scattering cross sections are linear functionals of the suitable parton distributions in the projectile and target \cite{Nikolaev:2005qs}. In \cite{Nikolaev:2003zf} a remarkable breaking of linear $k_T$-factorization was verified in forward dijet production in Deep Inelastic Scattering (DIS) off nuclei. Afterwards, this fact confirmed to be true in case of single-jet spectra in hadron-nucleus collisions \cite{Nikolaev:2004cu}. In \cite{Nikolaev:2005qs}, authors argued that dijet spectra and single-jet spectra in hadron-nucleus collisions clearly proved to be highly non linear of the collective nuclear gluon distribution. In addition, it was seen that the pattern of non linearity for single-jet spectra depends highly on the relevant partonic subprocesses \cite{Nikolaev:2003zf}. The breaking of linear $k_T$-factorization has been attested also in \cite{Blaizot:2004wv,JalilianMarian:2004da}. This means that
the color coupled channel aspect of the intranuclear color dipole evolution in general cannot be
absorbed into a single nuclear UGD. We are aware of this limitation and the calculation presented here is based on phenomenological arguments. For instance, the nuclear structure functions have been well described using factorization for nuclear targets in Refs.~\cite{Modarres:2019ndk,Modarres:2018ymh,Betemps:2010ay,deOliveira:2013oma}.

\section{Conclusions}

In this work we looked into timelike Compton scattering in nuclear processes. To do so, we applied the $k_T$-factorization approach and calculated the cross sections for dilepton production at the center-of-mass energies of different machines. Through our fit of TCS cross section in terms of the atomic mass number, it is possible to predict this observable for any nucleus. Predictions are make for $eA$ and ultraperipleral heavy ion collisions as well. 
Also, the investigation of nuclear TCS is quite relevant for constraining the GPDs and further analyses will certainly be highly important.

This is an exploratory study and is based on the fact that $k_T$-factorization is applicable for nuclei. In that sense, this study opens the possibility of investigating this issue. In the future, a comparison against data is in order to quantify the breaking of linear $k_T$ - factorization.

\label{summary}

\section*{Acknowledgments}

This work was financed by the Brazilian funding
agency CNPq. We are thankful to Emmanuel Gräve de Oliveira for his careful reading and highly important suggestions concerning this work.

\bibliographystyle{h-physrev}
\bibliography{referencias_tcs}

\end{document}